\documentclass[12pt,epsf,amssymb,qsymbols]{article}
\usepackage{tabularx}
\usepackage{array}
\usepackage{graphics}
\usepackage{graphicx}
\usepackage{psfrag}
\usepackage{epsfig}
\usepackage{amsmath}
\usepackage{amssymb}
\usepackage{ulem}
\usepackage{rotating}
\usepackage{colortbl}
\usepackage{tabularx}
\usepackage{longtable}
\usepackage{multirow}
\makeatletter

\usepackage{verbatim}

\setlongtables

\newcommand{\bea}{\begin{eqnarray}}
\newcommand{\eea}{\end{eqnarray}}
\newcommand{\beq}{\begin{equation}}
\newcommand{\eeq}{\end{equation}}
\newcommand{\ec}{\end{center}}
\newcommand{\bc}{\begin{center}}
\newcommand{\gev}{{\rm GeV}}

\newcommand{\pdir}{p\kern -5.2pt\raise 0.2ex\hbox {/}}

\newcommand{\vdir}{v\kern -5.75pt\raise 0.15ex\hbox {/}}
\newcommand{\kdir}{k\kern -5.75pt\raise 0.15ex\hbox {/}}
\newcommand{\epsdir}{\epsilon\kern -5.0pt\raise 0.15ex\hbox {/}}
\newcommand{\bvdir}{\bar{v}\kern -5.75pt\raise 0.15ex\hbox {/}}
\newcommand{\Ddir}{D\kern -7.75pt\raise 0.20ex\hbox {/}}
\newcommand{\Adir}{A\kern -7.75pt\raise 0.20ex\hbox {/}}
\newcommand{\ldir}{l\kern -5.0pt\raise 0.2ex\hbox{/}}
\newcommand{\varepsdir}{\varepsilon\kern -5.5pt\raise 0.15ex\hbox{/}}

\newcommand{\nn}{\nonumber}
\makeatother

\begin{document}
\thispagestyle{empty} 
\begin{flushright}
\begin{tabular}{l}
LPT 12-10
\end{tabular}
\end{flushright}
\begin{center}
\vskip 1.5cm\par
{\par\centering \textbf{\LARGE  
\Large \bf Pionic couplings to the lowest heavy-light mesons}}\\
\vskip 0.25cm\par
{\par\centering \textbf{\LARGE  
\Large \bf of positive and negative parity }}\\
\vskip 2.25cm\par
{\scalebox{.87}{\par\centering \large  
\sc Damir Be\v{c}irevi\'c, Emmanuel~Chang and Alain Le Yaouanc}}
{\par\centering \vskip 0.75 cm\par}
{\sl 
Laboratoire de Physique Th\'eorique (B\^at.~210)~\footnote{Laboratoire de Physique Th\'eorique est une unit\'e mixte de recherche du CNRS, UMR 8627.}\\
Universit\'e Paris Sud, Centre d'Orsay,\\ 
F-91405 Orsay-Cedex, France.}\\
{\par\centering \vskip 0.4 cm\par}
{\sl 
Dept. dÕEstructura i Constituents dela Mat\`eria, \\
Institut de Ci\`encies del Cosmos (ICC), \\
Universitat de Barcelona, Mart\`i Franqu\`es 1, E08028-Spain.}\\\vskip1.cm

{\vskip 0.35cm\par}
\end{center}

\vskip 0.55cm
\begin{abstract}
We present the method and compute the strong couplings of the lowest and first orbitally excited heavy-light mesons to a soft pion in the static heavy quark limit on the lattice. Besides the usual $\hat g$ and $\widetilde g$ couplings, we were able to make the first computation of the coupling $h$ using the relevant radial distributions. 
Our results are obtained from the simulations of QCD with $N_{\rm f}=2$ light Wilson-Clover quarks, combined with the improved static quark action.  
The hierarchy among couplings that emerges from our study is $\widetilde g < \hat g < h$. 
\end{abstract}
\vskip 2.2 cm 
\setcounter{page}{1}
\setcounter{footnote}{0}
\setcounter{equation}{0}
%%%%%%%%%%%%%%%%%%%%%%%%%%%%%%%%%%%%%%%%
\noindent

\renewcommand{\thefootnote}{\arabic{footnote}}

\newpage
\setcounter{footnote}{0}
%%%%%%%%%%%  Section 1
\section{Introduction}
A lattice QCD estimate of the pionic couplings to the pair of heavy-light mesons belonging to the same or different doublet(s) is important for at least three reasons: (i) They parameterize the hadronic matrix element relevant to the kinematically allowed $D^\ast \to D\pi$, $D^\ast_0\to D\pi$, $D^\prime_1\to D^\ast_0 \pi$, and the similar $B$-decay modes when they are kinematically allowed; (ii) They help us understanding better the viability of various  model-approaches used in describing the hadronic interactions at low energies, that are still used whenever the calculations based on simulations of QCD on the lattice cannot be made. Furthermore, these couplings are helpful in understanding the dynamics in crossed channels that are expected to drive the behavior of the desired $D\to \pi$ and $B\to \pi$ form factors; (iii) They enter the expressions derived in heavy meson chiral perturbation theory (HMChPT) that are used to guide the chiral extrapolations of the $B$-physics quantities computed on the lattice. 

The simplest such a quantity is the so called $\hat g$ coupling that essentially represents  the hadronic matrix element of  the light axial current between the meson states belonging to the $j^P_\ell = (1/2)^-$ doublet. It was computed in the static heavy quark limit on the lattice, first in the quenched approximation ($N_f=0$)~\cite{g-ukqcd, g-orsay}, and later on by including  $N_f=2$ dynamical quark flavors~\cite{g-jlqcd,ours1,g-alpha,g-detmold}. For the propagating charm quark this coupling has been computed in refs.~\cite{gDDpi1,gDDpi2} and it turned out to be considerably larger than in the static heavy quark limit.  Moreover,  the values for $\hat g$-coupling obtained in lattice QCD were much larger than the ones obtained by using various QCD sum rules techniques (c.f. ref.~\cite{dirac} and references therein) widely used in the $B$-physics phenomenology. Finally, the result obtained  with the propagating charm quark appeared to be consistent with the experimentally established vector meson width $\Gamma(D^{\ast})$~\cite{CLEO}.

Similar pionic couplings that relate an orbitally excited  and the lowest pseudo-scalar heavy-light mesons are harder to compute, especially in the static heavy quark limit. To our knowledge the only attempt in that direction has been made in ref.~\cite{michael}, where the authors computed the $B^\ast_0 \to B\pi$ decay amplitude in the static heavy quark limit directly, without relying on the soft pion theorem. 

We remind the reader that the heavy-light meson states belonging to the $j^P_\ell = (1/2)^-$ doublet are denoted as $(B, B^\ast)$ having $J^P=(0^-, 1^-)$, while the first orbital excitations $j^P_\ell   = (1/2)^+$, are denoted as $(B_0^\ast, B_1^\prime)$, and have quantum numbers $J^P=(0^+,1^+)$. A similar notation is used for the corresponding $D$-meson states.  

In our recent paper~\cite{ours2} we presented the results of our study of the radial distributions of the light quark axial current density sandwiched by the static meson states both belonging to either $j^P_\ell = (1/2)^-$ or $(1/2)^+$ doublets. That study was a continuation of the research made in ref.~\cite{green} in which we implemented the improved form of the Wilson line  on the lattice (static heavy quark propagator) by using the so-called hyper-cubic blocking (HYP), a procedure described in detail in refs.~\cite{hasenfratz,actions-dellamorte}. We showed that the accuracy of the lattice data can be further improved by applying the HYP blocking on the heavy spectator quark twice. Later on the radial distributions reported in ref.~\cite{ours2} were compared with those obtained in two classes of quark models and a striking qualitative and even quantitative agreement has been found~\cite{ours3}.

The integration of the radial distributions of the light axial current densities leads to the pionic couplings $\hat g$ and $\widetilde g$, for the case of heavy-light mesons belonging to $j^P_\ell = (1/2)^-$ and $(1/2)^+$, respectively.  The fact that we were able to study the radial distributions to a good accuracy opened a possibility for computing the axial coupling $h$, that relates the heavy light mesons belonging to two different doublets, such as the one that parameterizes the hadronic amplitude in $B^\ast_0 \to B\pi$ decay, i.e. the scalar to pseudoscalar decay with the emission of an $S$-wave pion. 

In the following we will describe how the coupling $h$ can be computed on the lattice in the static heavy quark limit. We will remind the reader of the difficulties  encountered in the study of $h$, and present our solution of the problem. We will then present the numerical results from our  lattice QCD calculations with $N_f=2$ dynamical quark flavors of Wilson type. In addition, we will provide the other couplings,  $\hat g$ and $\widetilde g$, that are then combined with $h$-coupling in the combined chiral extrapolation. 

\section{Methodology}
\subsection{Definitions and relevant hadronic matrix elements}
We begin by explaining the method to get the $h$-coupling in the soft pion limit. The starting point is the hadronic matrix element of the light quark axial current between the scalar ($B_0^\ast$), and the pseudoscalar ($B$) mesons,
\bea
\langle B(p^\prime)\vert A_\mu(0)\vert B_0^\ast(p)\rangle = (p+p^\prime)_\mu A_+(q^2) + q_\mu A_-(q^2)\,,
\eea
where $A_\pm(q^2)$ are two Lorentz invariant form factors, functions of the momentum transfer $q^2=(p-p^\prime)^2$, and the charged light axial current is ${A}_\mu = \bar u\gamma_\mu \gamma_5 d$, that in the following will be denoted as ${A}_\mu = \bar q\gamma_\mu \gamma_5 q$, reflecting the fact that  we take the quarks to be degenerate in mass, $m_u=m_d\equiv m_q$. After taking the divergence of the axial current in the soft pion limit we have,
\bea
\lim_{q^2\to 0}\langle B(p^\prime)\vert q^\mu A_\mu(0)\vert B_0^\ast(p)\rangle =  (m_{B_0^\ast}^2 - m_B^2) A_+(0)\, .
\eea
On the other hand, by means of the reduction formula,
\bea
{f_\pi m_\pi^2\over m_\pi^2-q^2}  \langle \pi^\pm(q) B(p^\prime)\vert B_0^\ast(p)\rangle  = \langle B(p^\prime)\vert q^\mu A_\mu(0)\vert B_0^\ast(p)\rangle \,,
\eea
so that by defining 
\bea
\langle \pi^\pm(q) B(p^\prime)\vert B_0^\ast(p)\rangle = g_{B_0^{\ast }B\pi}\,,
\eea
in the same  limit,  $q^2\to 0$, we have
\bea
  g_{B_0^{\ast }B\pi} =  {  m_{B_0^\ast}^2 - m_B^2\over f_\pi } A_+(0)\,.
\eea
The definition which relates the coupling $g_{B_0^{\ast }B\pi}$ with the coupling $h$, that appear in the chiral Lagrangian~\cite{Casalbuoni}, reads
\bea
g_{B_0^{\ast }B\pi} =\sqrt{m_B m_{B_0^{\ast }} }    { m_{B_0^{\ast }}^2 - m_B^2 \over m_{B_0^{\ast }} }  { h\over f_\pi}\,,
\eea
and therefore in the static heavy quark mass limit we can identify,  
\bea
h=A_+(0)\,.
\eea
This coupling describes the emission of the $S$-wave soft pion off the $B_0^\ast$ meson, i.e. 
\bea\label{eq:xxx}
&&\Gamma(B_0^\ast\to B \pi^\pm)={g_{B^{\ast}_0 B\pi}^2\over 8 \pi m_{B_0^\ast}^2} |\vec q_\pi|\,,\hfill \\
&&|\vec q_\pi|={\sqrt{[m_{B_0^{\ast }}^2 - (m_B-m_\pi)^2]  [m_{B_0^{\ast }}^2 - (m_B+ m_\pi)^2] } \over 2 m_{B_0^{\ast }} 
},
\eea
and similarly for the charmed heavy-light mesons. 

The main goal of the present paper is therefore the extraction of the form factor $A_+(0)$. 
To ensure the standard heavy quark effective theory (HQET) normalization of states,  $ \langle B_a(v)\vert B_b(v^\prime)\rangle = \delta_{ab}\delta(v - v^\prime)$, 
and by taking both {\it in} and {\it out} states to be at rest, we rewrite
\bea
   \langle B \vert A_0 (0)\vert B_0^\ast \rangle^{\rm HQET} = {m_B + m_{B_0^{\ast }} \over 2\sqrt{m_B m_{B_0^{\ast }} } } A_+(\Delta^2)  
 + {\Delta \over 2\sqrt{m_B m_{B_0^{\ast }} } } A_-(\Delta^2)\,,
\eea
where we introduced $\Delta =  m_{B_0^{\ast }} - m_B$, the mass difference between the lowest orbitally excited heavy-light meson and its $L=0$ counterpart. In the static heavy quark limit ($m_b\to \infty$)  we finally have,
\bea
 \langle B \vert A_0 (0)\vert B_0^\ast \rangle^{\rm HQET} &=& A_+(\Delta^2)\,.
\eea 
Notice that the argument of the form factor is non-zero even in the static heavy quark limit ($\Delta \neq 0$) because the two mesons belong to different doublets of heavy-light meson states. 
In order to go from $A_+(\Delta^2)$ to the coupling $h=A_+(0)$, an extra step is needed. Before we discuss that issue, we will spend some time explaining how the $A_+(\Delta^2)$ is computed on the lattice.
\subsection{Extraction of $A_+(\Delta^2)$ from the correlation functions computed on the lattice}
On the lattice we first compute the following three-point correlation functions
\bea\label{r-1}
C_{PAS}(t_y,t_x)&=&\langle\sum_{\vec x,\vec y} P(y) {A}_0(0) S^\dagger(x)\rangle_{_U}, \cr
C_{SAP}(t_y,t_x)&=&\langle\sum_{\vec x,\vec y} S(y) {A}_0(0) P^\dagger(x)\rangle_{_U}, \cr
C_{PVP}(t_y,t_x)&=&\langle\sum_{\vec x,\vec y} P(y) {V}_0(0) P^\dagger(x)\rangle_{_U}, \cr
C_{SVS}(t_y,t_x)&=&\langle\sum_{\vec x,\vec y} S(y) {V}_0(0) S^\dagger(x)\rangle_{_U} ,
\eea
where $\langle\dots\rangle_{_U}$ denotes the average over independent gauge field configurations, the interpolating operators for the pseudoscalar and scalar heavy-light mesons are respectively $P= \bar h \gamma_5 q$ and $S= \bar h \mathbb{I} q$, with $h(x)$ and $q(x)$ being the static heavy and the light quark fields respectively. 
The purpose of computing the elastic correlation functions, $C_{PVP,SVS}(t_y,t_x)$ with ${V}_0=\bar q\gamma_0 q$, is to eliminate the interpolating source operators. Those elastic matrix elements are fixed by the electric charge conservation and,  after accounting for the appropriate renormalization constant,  the relevant form factors are simply equal to one. To extract $A_+(\Delta^2)$ 
we adopt the strategy of double-ratios and compute 
\bea\label{double}
 R_h(t)&=&-{Z_A^2\over Z_V^2}\  {C_{SAP}(t_y,t_x)\ C_{PAS}(t_y,t_x)\over C_{PVP}(t_y,t_x)\ C_{SVS}(t_y,t_x)} \nn\\
 && \hfill \nn \\
&&\hspace*{-5mm} \xrightarrow[]{\hspace*{5mm} } -{ 
\langle B_0^\ast \vert A_0  \vert B \rangle\,  \langle B \vert A_0 \vert B_0^\ast \rangle 
   \over  \langle B \vert V_0  \vert B \rangle \,  \langle B_0^\ast \vert V_0  \vert B_0^\ast  \rangle 
   }  = [A_+(\Delta^2)]^2\,,
\eea
where an extra ``$-$" sign accounts for the fact that the matrix elements for the pion emission and the pion absorption have opposite signs.  In terms of quark propagators 
\bea\label{eq:correlators}
C_{PAS}(t_y,t_x) &=& \langle\sum_{\vec x,\vec y} {\rm Tr}\left[ {1+\gamma_0\over 2}W_x^y \gamma_5{\cal S}_q(y,0) \gamma_0\gamma_5
{\cal S}_q(0,x)\mathbb{I}\right]\rangle_{_U}   \cr
& =&  -\langle\sum_{\vec x} {\rm Tr}\left[ {1+\gamma_0\over 2}V_x^y  {\cal S}^\dagger_q(\vec 0,0;\vec x,t_y) \gamma_0
{\cal S}_q(\vec 0,0 ;\vec x,t_x ) \right]\rangle_{_U} ,
\eea
where we used the $\gamma_5$-hermiticity of the quark propagator in the euclidean space, ${\cal S}^\dagger_q(x,y)= \gamma_5 {\cal S}_q(y,x)\gamma_5$. In addition, we wrote the static quark propagator as 
$(1+\gamma_0)/2 \times W_{x}^y$, with the Wilson line being
\bea\label{Pline}
W_x^y = \delta(\vec x-\vec y) \prod_{\tau=t_y}^{t_x-1}U^{\rm impr.}_0(\tau, \vec x) \equiv \delta(\vec x-\vec y) V_x^y \,.
\eea
The latter is merely obtained from the discretized static heavy quark action~\cite{Eichten-Hill}
\bea \label{hqet-lagr}
{\cal L}_{\rm HQET}=\sum_x  h^\dagger(x)\left[ h(x) - U^{\rm impr.}_0(x-\hat 0)^\dagger h(x-\hat 0)\right] \,,
\eea
where for  the time component of the link variable, $U^{\rm impr.}_0$, we use several improved schemes discussed in our previous paper~\cite{ours2}.  
Improvement of the discretized Wilson line is essential as it ensures the exponential improvement of the signal to noise ratio in the correlation functions with respect to 
what is obtained by using the simple product of link variables~\cite{alphaSN}.
Similarly, other correlation functions are computed as
\bea\label{eq:correlators2}
C_{SAP}(t_y,t_x) &=& \langle\sum_{\vec x} {\rm Tr}\left[ {1+\gamma_0\over 2}V_x^y {\cal S}_q(\vec x,t_y ;\vec 0,0 ) \gamma_0
{\cal S}^\dagger_q(\vec x,t_x;\vec 0,0  ) \right]\rangle_{_U} \,,\cr
C_{PVP}(t_y,t_x) &=& \langle\sum_{\vec x} {\rm Tr}\left[ {1+\gamma_0\over 2}V_x^y {\cal S}^\dagger_q( \vec 0,0 ; \vec x,t_y ) \gamma_5 \gamma_0
 {\cal S}_q( \vec 0,0 ; \vec x,t_x ) \gamma_5 \right]\rangle_{_U} \,,\cr
C_{SVS}(t_y,t_x) &=& \langle\sum_{\vec x} {\rm Tr}\left[ {1-\gamma_0\over 2}V_x^y {\cal S}^\dagger_q( \vec 0,0 ; \vec x,t_y ) \gamma_5 \gamma_0
 {\cal S}_q( \vec 0,0 ; \vec x,t_x ) \gamma_5 \right]\rangle_{_U} \,.
\eea
In the following we drop the dependence on $t_y$ as this time is kept fixed to one or several values that are sufficiently large so that the lowest lying state would be isolated. 
In practice we actually check that by reversing the roles of $t_x$ and $t_y$ our results remain stable. The use of improved static heavy quark actions is essential because we also need the signal 
to be good enough at large time separations between the source operators.  However some efficient smearing of the sources is welcome in order to have both 
sources sufficiently far away from the light quark current inserted between them, and therefore that the desired matrix element can be isolated. We use the smearing  proposed in ref.~\cite{Boyle},  
that we already discussed in our previous papers~\cite{ours1}, which consists in replacing the source operators  $\bar h(x) \Gamma  q(x)\to \bar h(x) \Gamma q^S(x)$, where $\Gamma$ is a Dirac matrix, and
\begin{align} \label{Smearing}
q^S(x)= \sum\limits_{r=0}^{R_{\rm max}} \varphi(r)
 \sum\limits_{k=x,y,z} \left[
q({  x+r\hat{k}})\times \prod\limits_{i=1}^r U_k({ 
x+(i-1)\hat{k}})  + q({  x-r \hat{k}})\times
\prod\limits_{i=1}^r U_k^{\dagger}({ 
x-i\hat{k}})\right]  \,,
\end{align} 
with $\varphi(r)=  e^{-r/R}(r+1/2)^2$. After inspection we find the smearing to be efficient  for  $R=3$ and $R_{\rm max}=5$.  
To make sure that the values we obtain are correct, we also tried the standard method of extracting the form factors from the three point functions, by dividing out the source operators obtained from the fit to two-point correlation functions computed on the lattice, i.e.,
\bea\label{eq:2pts}
&& C_{PP} (t_x) =
\sum_{\vec x}  \left<  \bar h(x)\gamma_5  q (x) \  \left(\bar h(0)\gamma_5   q (0)\right)^\dagger \right>_{_U} \, \to \, \sum_{i=1,2}{\cal Z}_{q i}^2 {\rm e}^{-{\cal E}_{qi} t_x} ,\cr
&&\nn\\
&&C_{SS} (t_x) =
\sum_{\vec x}  \left<  \bar h(x)\mathbb{I}  q (x) \  \left(\bar h(0)\mathbb{I}   q (0)\right)^\dagger \right>_{_U} \, \to \, \sum_{i=1,2}\widetilde {\cal Z}_{q i}^2 {\rm e}^{-\widetilde {\cal E}_{qi} t_x} .
\eea
We made that exercise by including either one or two exponentials on the right hand side in order to check the stability of our results. With all these checks we were able to find the stability window in time, that indeed leaves the form factor $A_+(\Delta^2)$ unchanged.  

\subsection{\label{q20}Going to $q^2\to 0$}
\begin{figure}
\begin{center}\includegraphics[width=11.5cm,clip]{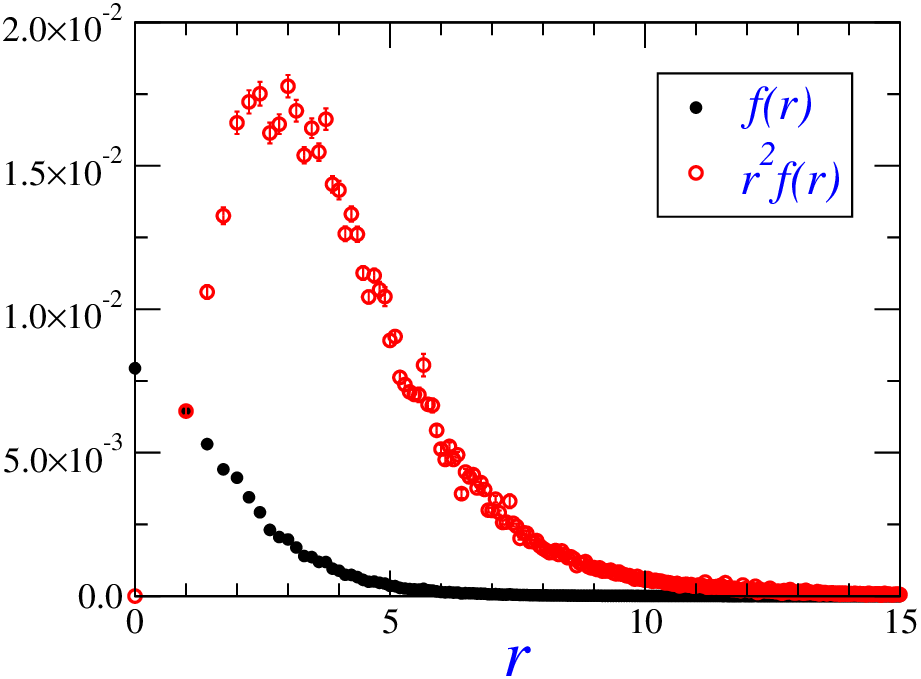}\end{center}
\caption{\label{fig:1} \footnotesize \sl Shapes of the distribution $f_{PAS}(r)$, as extracted from the three and two point correlation functions computed on the lattice as indicated in eq.~(\ref{fPAS}). Illustration is provided with one value of the sea quark mass corresponding to $\kappa_{\rm sea}=0.1374$ from Set-1 lattices listed in tab.~\ref{tab:1}, and with the HYP$^2$ static heavy quark propagator. All data points are given in lattice units.}
\end{figure}

As we already mentioned, the information we obtain from the study of the three point correlation functions with both source operators at rest is the form factor $A_+(\Delta^2)$, while we actually need $A_+(0)$. Going to $q^2=0$ would require a momentum injection to one of the source operators, which is particularly complicated to do in the case of static heavy-light mesons. Furthermore, tuning the three-momentum to $\vert \vec p-\vec p^\prime\vert = \Delta$ is very difficult too. This is where our experience with radial distributions of various light quark current matrix elements becomes useful. We follow the same procedure employed in ref.~\cite{ours2} and compute the radial distribution of  the correlation function $C_{PAS}$, namely
\bea
C_{PAS}(t_y,t_x; \vec r ) = \langle\sum_{\vec x,\vec y} \bar h(x)\gamma_5 q(x) ~~ \bar q(0+\vec r)\gamma_0\gamma_5 q(0+\vec r) ~~ (\bar h (y) \mathbb{I} q(y) )^\dagger \rangle_{_U} ,
\eea
that leads to
\bea\label{fPAS}
{ C_{PAS}(t_y,t_x; |\vec r| ) \over {\cal Z}^S_q \widetilde {\cal Z}^S_q \times \exp[ \widetilde {\cal E}_q t_y  + {\cal E}_q t_x]} \, \xrightarrow[]{\quad{-t_x, t_y\gg 0}\quad} \, \langle B_q \vert A_0(r)\vert B_0^\ast\rangle \equiv f_{PAS}(r)\,,
\eea
for the sufficiently large time separations among operators. Note that we place one source operator at $t_x<0$ and the other at $t_y>0$, while the light quark axial current, $A_0=\bar q\gamma_0\gamma_5 q$, is fixed at the origin of the lattice. The shape of the function $ f_{PAS}(r)$ for one representative value of the light quark mass is shown in fig.~\ref{fig:1}.~\footnote{Note that in our study we always consider the sea and the valence light quarks to be mass degenerate. } 

We checked that the integral,
\bea\label{integral0}
A_+(\Delta^2) = 4 \pi \int_0^\infty dr\ r^2  f_{PAS}(r)\,,
\eea
indeed reproduces the form factor value obtained from the double ratio~(\ref{double}), even if with slightly larger error bars.  
To get the form factor at $q^2\neq \Delta^2$  we then give the momentum $\vec q=(0,0,q_z)$, by taking the Fourier transform, 
\bea\label{eq:8}
 A_+(\Delta^2- q_z^2) &=& \int d\vec r  \ f_{PAS}(r) \ e^{i \vec q \vec r}= 2\pi  \int_0^\infty dr\ r^2  f_{PAS}(r) \int_0^\pi e^{i q_z r \cos \theta} \sin\theta d\theta \nn\\
&= &4\pi  \int_0^\infty dr\ r^2  {\sin q_zr\over q_zr} f_{PAS}(r)\,.
\eea
Obviously, by taking $q_z=\Delta \equiv \widetilde {\cal E}- {\cal E}$, that we easily extract from the study of the time dependence of two point correlation functions, as indicated in eq.~(\ref{eq:2pts}), we get the desired form factor $A_+(0)$, namely
\bea\label{eq:81}
A_+(0) = 4\pi  \int_0^\infty dr\ r^2  {\sin( \Delta r )\over \Delta  r} f_{PAS}(r)\,.
\eea
Since the mass dependence in renormalization constants with improved Wilson action on the lattice cancel to a large extent in the double ratio~(\ref{double}), we prefer to determine the  form factor $A_+(\Delta)$ in that way. Indeed, from the above eqs.~(\ref{eq:8},\ref{eq:81}), we can get the correction, 
\bea
R_\Delta = {A_+(0)  \over A_+(\Delta^2)} = {  {\displaystyle\int_0^\infty dr\ r^2 {\sin( \Delta r )\over \Delta  r}  f_{PAS}(r) } \over  \displaystyle{\int_0^\infty dr\ r^2  f_{PAS}(r)}}\,, 
\eea
and therefore our value for the desired pionic coupling $h$ will be obtained via, 
\bea
h_q = A_+(\Delta_q^2)\times  R_{\Delta_q} \,,
\eea
where in the last line we added an index $q$ to distinguish $h_q$ from the true coupling $h$ that is obtained in the chiral limit.

\subsection{$\hat g$ and $\widetilde g$ couplings on the lattice}
For a consistent chiral extrapolation of $h_q$, we also need the couplings $\hat g$ and $\widetilde g$, that we discussed in our previous papers~\cite{ours1,ours2}. With respect to ref.~\cite{ours2}, here we take the light quark current to be at the origin of the lattice, which helps avoiding  the fact that the ratio of correlation functions becomes a sum of all possible diagonal  matrix elements. Furthermore we extract the couplings from the ratios of three point correlation functions instead of three to two point functions. In short, the couplings $\hat g$ and $\widetilde g$ are defined via 
\bea
\langle B(\vec 0)\vert \bar q \gamma_i \gamma_5q\vert B^\ast(\varepsilon,\vec 0) \rangle =\varepsilon_i \ \hat g\,,\cr 
&&\hfill \nn\\
\langle B_0^\ast(\vec 0)\vert \bar q \gamma_i \gamma_5q\vert B_1^\prime(\varepsilon^\prime,\vec 0) \rangle = \varepsilon_i^\prime\ \widetilde g\,,
\eea
where $\varepsilon_\mu$ and $\varepsilon_\mu^\prime$ are the polarization vectors of the vector and axial-vector heavy-light meson, respectively. 
To extract the above couplings from the ratios of three-point correlation functions we also computed:
\bea
C_{PAV}(t_y,t_x)&=&\langle\sum_{i,\vec x,\vec y} P(y) {A}_0(0) V_i^\dagger(x)\rangle_{_U}, \cr
C_{SAA}(t_y,t_x)&=&\langle\sum_{i,\vec x,\vec y} S(y) {A}_i(0) A_i^\dagger(x)\rangle_{_U}, 
\eea
which when combined with those listed in eq.~(\ref{r-1}) lead to
\bea\label{gg2}
&&R_g(t_y,t_x)= {Z_A\over Z_V} \ {C_{PAV}(t_y,t_x)\over C_{PVP}(t_y,t_x)} \to
 {\, \langle B \vert A_i  \vert B^\ast \rangle\,  \over  \langle B \vert V_0  \vert B \rangle  } 
=\hat g_q\,,\nn\\
&&\widetilde R_g(t_y,t_x)= {Z_A\over Z_V} \ {C_{SAA}(t_y,t_x)\over C_{SVS}(t_y,t_x)} \to
 {\, \langle B_0^\ast \vert A_i  \vert B_1^\prime \rangle\,  \over  \langle B_0^\ast \vert V_0  \vert B_0^\ast \rangle  } 
=\widetilde g_q\,,
\eea
where we again added an index $q$ to distinguish  $\hat g_q$ and $\widetilde g_q$ from the true pionic coupling  that are defined in the soft pion limit. 
To our knowledge the computation of these couplings from the ratios of three-point correlation functions alone has not been attempted elsewhere.

\section{Lattice computation of the pionic couplings}

In this section we present our numerical results. We use three sets of publicly available gauge field configurations, all obtained by using the Wilson actions. In particular, the first set of data is obtained by using the Iwasaki gauge action, while for the other two sets the standard Wilson plaquette action has been used. As for the discretized Dirac part of the QCD action, in all three sets of configurations the Wilson-Clover action was implemented to ensure the elimination of ${\cal O}(a)$ discretization effects from the lattice action. More specifically, in the first set the Clover coefficient has been set to its value obtained by using the one-loop boosted perturbation theory ($c_{SW}=1.47$), while in the other two ensembles the non-perturbatively determined Clover coefficient was used in simulations ($c_{SW}=1.919$, and $1.823$, for $\beta=5.29$ and $5.40$, respectively). Obviously, in computation of the light quark propagators we used the same values of $c_{SW}$ used to generate the gauge field configurations with $N_{\rm f}=2$ flavors of mass degenerate light quarks. From the set of publicly available gauge field configurations we chose those that are separated by $20$ unit-length HMC trajectories. 

More details concerning the simulations from which the gauge field configurations used in this work have been obtained can be found in refs.~\cite{cp-pacs, qcdsf}. In tab.~\ref{tab:1} we summarize the main details concerning the gauge ensembles used in this work.  
\begin{table}[t]
\begin{center}
\begin{tabular}{|c|cccc|cc|}\hline
{\phantom{\Large{l}}} \raisebox{-.1cm} {\phantom{\Large{j}}}
\hspace*{-3.5mm} Action &$ \beta$ [$r_0/a$] &  $\kappa_{\rm sea}$ & $m_\pi$ [GeV]& ref. & $Z$ (NP) & $Z$  (TI) \\ \hline\hline
{\phantom{\Large{l}}}  {\phantom{\Large{j}}}
\hspace*{-5.5mm} Iwasaki/Clover & $2.1$ [$4.70(2)$]    &   0.1357   & 1.025(5) & \cite{cp-pacs}&1.03(2) &	1.017	\\
\hspace*{-3.5mm}{\color{blue}Set 1}   &       &   0.1367  &  0.888(4)&   & &	\\
\hspace*{-3.5mm}   &      &    0.1374   & 0.761(4)&  & & 	\\
\hspace*{-3.5mm}   &      &    0.1382  &   0.563(4) & 	& & \\ \hline
{\phantom{\Large{l}}}   {\phantom{\Large{j}}}
\hspace*{-3.5mm}WP/Clover & $5.29$ [$6.20(3)$]       &    0.1355  &   0.858(6) & \cite{qcdsf}& 1.038 &  1.029	\\
\hspace*{-3.5mm} {\color{blue}Set 2}  &      &    0.1359   &   0.643(7) & 	& &\\
\hspace*{-3.5mm}   &       & 0.1362  &   0.406(6)& &	& \\  \hline
{\phantom{\Large{l}}}   {\phantom{\Large{j}}}
\hspace*{-3.5mm}WP/Clover & $5.40$ [$6.95(4)$]      &  0.1356   &   0.969(6)  & \cite{qcdsf}&1.033 &	1.028 \\
\hspace*{-3.5mm} {\color{blue}Set 3}  &       &  0.1361  &  0.672(7) &  & &	\\
\hspace*{-3.5mm}   &       &  0.13625  &  0.572(8) &  &	  & \\  \hline
\end{tabular}\caption{\label{tab:1}\footnotesize  \sl Gauge field configurations with $N_f=2$ dynamical Wilson-Clover quarks, 
where the hopping parameters  $\kappa_{\rm sea}$ is specified together with the corresponding pion mass. ``WP" stands for the Wilson-Plaquette gauge action, and Clover stands for the ${\cal O}(a)$-improved Wilson quark action. 
All lattice volumes are $24^3\times 48$. More information on each set of configurations can be found in the quoted references.  The ratio of renormalization constants is denoted by $Z=Z_A(g_0^2)/Z_V(g_0^2)$. ``NP" stands for non-perturbatively determined value, that is compared to a tadpole improved perturbative one (``TI").}
\end{center}
\vspace*{-3mm}\end{table} 
To situate the values of the lattice spacings, one can use $r_0=0.47$~fm~\cite{r047}, which then translates to $a^{-1}\simeq 2.0$, $2.6$ and $3.1$~GeV, for our three sets of gauge field configurations respectively. It should be stressed that the value of $r_0$ is unphysical and is still a subject of uncertainty. For example, from simulations with twisted mass QCD on the lattice and by using the physical pion decay constant one obtains $r_0=0.44(1)$~fm~\cite{Blossier:2009bx}. For our final results, the uncertainty in $r_0$ is unimportant as our main goal is to obtain the dimensionless pionic couplings in the chiral limit.

In tab.~\ref{tab:1} we also give the values of $Z=Z_A(g_0^2)/Z_V(g_0^2)$ that is used to extract $A_+(\Delta_q)$, $\hat g$, and $\widetilde g$, from $R_h$, $R_g$, and $\widetilde R_g$ respectively. Besides the cancellation of the exponential terms these ratios are convenient because the sizable mass corrections to the renormalization constants cancel to a large extent in the ratio. More specifically, for the consistent ${\cal O}(a)$-improvement of the Wilson quark action, the renormalization constants are improved as: 
\bea
Z_{V,A}(g_0^2)= Z_{V,A}^{(0)}(g_0^2) \left[ 1 + b_{V,A}(g_0^2) (a m_q) \right]\,,
\eea
where $Z_{V,A}^{(0)}(g_0^2)$ stands for the value obtained in the chiral limit, and $b_{V,A}(g_0^2)$ is the counter-term coefficient chosen to eliminate  the   
 the ${\cal O}(am_q)$ effects when working with the vector and axial-vector current respectively. 
 Knowing that $b_{V}\approx b_{A}$, the effect of  ${\cal O}(am_q)$ corrections to $Z=Z_A(g_0^2)/Z_V(g_0^2)$ becomes negligible.  Furthermore the non-perturbatively determined value of $Z$ in the chiral limit has been found to be very close to its value predicted by using the tadpole improved one-loop perturbation theory~\cite{TI}. This can be appreciated from the numbers in tab.~\ref{tab:1} where the non-perturbatively computed $Z$~\cite{Gockeler:2010yr} are less than $1\%$ larger than the corresponding tadpole improved perturbative estimate~\cite{TI,Gockeler:2010yr}. In the first set instead the non-perturbative value for the ratio was deduced from the Ward identities~\cite{ours1}, which turns out to be very close to the perturbative result~\cite{cp-pacs}. In what follows, for the first set we will  use the perturbative result (TI), while for the other two  the non-perturbative values will be used. 

Our main results will be obtained from Set 1 of the lattice data for which we were able to compute all three couplings. With the other two ensembles the computation of $\widetilde g_q$ was unstable, but the computation of $\hat g_q$ and $h_q$ is good which is why we prefer to separate the discussion of our results. 

\subsection{Discussion of the numerical results}

\begin{figure}
\psfrag{AA}{\Large${\color{blue}({\cal E}_{q} + \widetilde {\cal E}_{q})(t)}$}
\psfrag{BB}{\Large${\color{blue}\widetilde {\cal E}_{q}^{(2)}(t)}$}
\psfrag{CC}{\Large${\color{blue} {\cal E}_{q}^{(2)}(t)}$}
\begin{center}\includegraphics[width=11.1cm,clip]{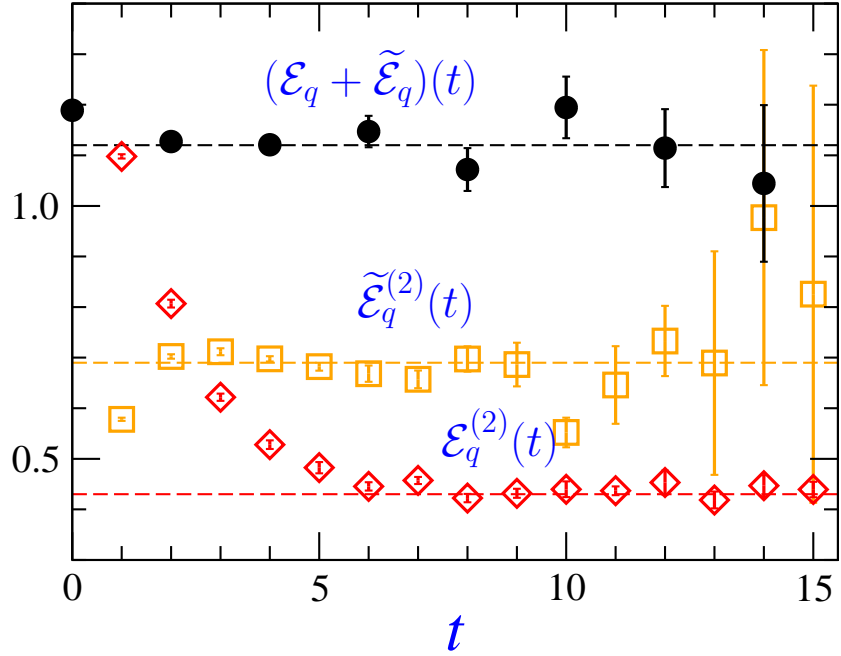}\end{center}
\caption{\label{fig:2} \footnotesize \sl Effective mass plots: ${\cal E}_{q}^{(2)}(t)$ and $\widetilde {\cal E}_{q}^{(2)}(t)$ refer to the $(1/2)^-$ and $(1/2)^+$ states respectively, both obtained from the two-point correlation functions computed on the lattice as indicated in eq.~(\ref{eff-2pts}); $({\cal E}_{q} + \widetilde {\cal E}_{q})(t)$ is obtained from the three-point correlation functions, c.f. eq.~(\ref{eff-3pts}). Dashed lines correspond to the central values of the static heavy-light binding energy:  ${\cal E}_{q}^{(2)}$, $\widetilde {\cal E}_{q}^{(2)}$, and ${\cal E}_{q}^{(2)}+\widetilde {\cal E}_{q}^{(2)}$. Illustration is provided for the lattices of Set 1, with $\kappa_{\rm sea}\equiv \kappa_q=0.1374$, and by using the HYP-2$^2$ static heavy quark propagator. All data points are given in lattice units.}
\end{figure}

We summarize our numerical results in tab.~\ref{tab:2} that we comment in the following. The mass difference between the lowest lying  $(1/2)^+$ and $(1/2)^-$ states, $\Delta_q = \widetilde {\cal E}_q - {\cal E}_q$, is obtained from the fit of the form given in eq.~(\ref{eq:2pts}) to the two-point correlation functions computed on the lattice. All results presented in this section are obtained by using the HYP-2$^{2}$ static quark action discussed in our previous paper~\cite{ours2}. With other forms of the static quark actions we obtain fully compatible results with those listed in tab.~\ref{tab:2}.  
\begin{table}[t!!]
\begin{tabular}{|c|c|c|c|c|c|c|} 
\hline
{\phantom{\huge{l}}} \raisebox{-.2cm} {\phantom{\huge{j}}}
 $\kappa_{q}\hspace*{5mm}$ &$\hspace*{6mm} \Delta_q\hspace*{6mm}$&  $\hspace*{6mm}\hat g_q\hspace*{6mm}$  & $\hspace*{6mm}\widetilde g_q\hspace*{6mm}$ & $\hspace*{2mm}A_+(\Delta_q)\hspace*{6mm}$ & $R_{\Delta_q}$ & $h_q=A_+(0)$ \\ \hline \hline
{\phantom{\huge{l}}} \raisebox{-.2cm} {\phantom{\huge{j}}}
                                                 $0.1357$     & $0.269(4)$  & $0.639(5)$ & $-0.040(10)$ & $0.809(24)$	 & $0.889(26)$ & $0.719(29)$ \\
{\phantom{\huge{l}}} \raisebox{-.2cm} {\phantom{\huge{j}}}
                                                 $0.1367$    & $0.273(4)$  & $0.621(3)$ & $-0.068(6)$ & $0.832(14)$& $0.896(15)$ & $0.746(18)$	\\
{\phantom{\huge{l}}} \raisebox{-.2cm} {\phantom{\huge{j}}}
                                                $0.1374$     & $0.262(6)$  & $0.579(9)$ & $-0.103(12)$ & $0.889(26)$& $0.842(24)$ & $0.751(30)$	\\
{\phantom{\huge{l}}} \raisebox{-.2cm} {\phantom{\huge{j}}}
                                                $0.1382$    & $0.272(6)$  & $0.529(19)$ & $-0.171(19)$ & $0.968(39)$& $0.808(32)$ & $0.783(44)$	\\ \hline
                                                \hline
{\phantom{\huge{l}}} \raisebox{-.2cm} {\phantom{\huge{j}}}
                                               $0.1355$ & $0.232(8)$      & $0.552(19)$ & $-0.094(10)$ & $0.981(38)$ &  $0.829(32)$ &	$0.814(45)$\\
{\phantom{\huge{l}}} \raisebox{-.2cm} {\phantom{\huge{j}}}
   			                 $0.1359$ & $0.210(5)$    & $0.514(17)$ & $-0.143(17)$ & $0.892(31)$	&  $0.797(28)$&  $0.711(35)$\\
{\phantom{\huge{l}}} \raisebox{-.2cm} {\phantom{\huge{j}}}
                                                $0.1362$ &$0.223(12)$    & $0.454(23)$ & $-0.171(20)$ & $0.919(56)$	&$0.772(47)$   &   $0.710(61)$  \\
	 \hline \hline
{\phantom{\huge{l}}} \raisebox{-.2cm} {\phantom{\huge{j}}}
 		                            $0.1356$ & $0.216(6)$     & $0.573(15)$ & $-0.110(11)$ & $0.978(40)$&  $0.840(35)$ &  $0.822(48)$	\\
{\phantom{\huge{l}}} \raisebox{-.2cm} {\phantom{\huge{j}}}
 					 $0.1361$ & $0.233(7)$    & $0.519(24)$ & $-0.155(11)$ & $0.903(32)$	&  $0.822(29)$ &  $0.743(37)$  \\
{\phantom{\huge{l}}} \raisebox{-.2cm} {\phantom{\huge{j}}}
           	                               $0.13625$ & $0.207(8)$    & $0.483(19)$ & $-0.164(8)$ & $0.868(23)$& $0.810(21)$ & $0.703(26)$	\\
	 \hline 

\end{tabular}\caption{\label{tab:2}\footnotesize Direct numerical results extracted from the correlation functions calculated on all of the ensembles of the lattices 
with parameters listed in tab.~\ref{tab:2}. Results are presented in the same order as in tab.~\ref{tab:2}.}
\end{table}
The benefit of using the improved static quark actions is that the signal remains good at larger time separations so that we could check that with either local or smeared sources we could obtain results that are compatible. Importantly the smearing helps to isolate the lowest lying state at shorter time separation between the source operators which is important for the studies of the three point functions. We make several tests by fitting each of the three point functions to its hadronic decomposition, 
\bea
{ C_{PAS}(t_y,t_x ) }= \sum_{i,j} ({\cal Z}^S_q)_i \, e^{ {\cal E}_{q i}\ t_x} \times   \langle (B_q)_i \vert A_0(0)\vert (B_0^\ast)_j \rangle \times ( \widetilde {\cal Z}^S_q)_j \, e^{  \widetilde {\cal E}_{q j} t_y  }\,,
\eea
with either $i,j=1$ or $i,j=2$, where the coupling to smeared source operators $({\cal Z}^S_q)_i$  and  $( \widetilde {\cal Z}^S_q)_j$ are obtained from the fits to two point correlation functions. In this way we could determine the stability interval where the fit to our data lead to a stable value of $A_+(\Delta_q^2)$. That is then compared with what we obtain by using the double ratio. Another good test for the goodness of the fitting interval is to compare the effective mass plots. We fix $-t_x=t_y=t$ and since the light quark axial current is fixed at the origin of the lattice, the effective mass from three point functions $C_{PAS}(t,-t)\equiv C_{PAS}(t)$:
\bea\label{eff-3pts}
({\cal E}_{q} + \widetilde {\cal E}_{q})(t) = \log\left[ {C_{PAS}(t+ 1) \over C_{PAS} (t )}\right]\,,
\eea
should be compatible with the sum of what we obtain from the two point functions 
\bea\label{eff-2pts}
&&{\cal E}_{q}^{\rm (2)}(t_x) = \log\left[ {C_{PP} (t_x) \over C_{PP} (t_x+1)}\right]\,,\qquad
\widetilde {\cal E}_{q}^{\rm (2)}(t_x) = \log\left[{ C_{SS} (t_x) \over C_{SS} (t_x+1)}\right]\,.
\eea
Since both interpolating field operators are pulled away from each other in eq.~(\ref{eff-3pts}) by one time unit, that makes twice larger distances of the source operators in time, than it is the case with two point correlation functions obtained from eq.~(\ref{eff-2pts}) where one source operator is kept fixed at $t_x=0$. After accounting for that fact we checked that for each of our lattices and for every three point correlation functions, the effective mass of three point functions is consistent with what is obtained from the two point functions. An illustration is provided in fig.~\ref{fig:2}, which is a typical situation with the lattice data. As expected the signal for the excited [$(1/2)^+$] state is less good than the one for the ground static heavy-light state  [$(1/2)^-$]. 
The values of $A_+(\Delta_q)$, $\hat g_q$, and $\widetilde g_q$  are then obtained from the ratios (\ref{double}) and  (\ref{gg2}). As we mentioned above, besides the double ratios we also used the ratios with two point correlation functions similar to  the one in eq.~(\ref{fPAS}) to check for the consistency of the obtained results. This also helped us determine the sign of the form factor $A_+(\Delta_q)$. After a detailed inspection, we fit our data between  $-t_x=t_y=t\in [4,7]$ for $A_+(\Delta_q)$, between $t\in [3,9]$ for $\hat g_q$, and $t\in [3,5]$ for $\widetilde g_q$. Finally, we should emphasize once again that all the results presented here are fully unquenched, i.e. the valence quark mass is equal to the sea quark mass.

\begin{figure}
\begin{center}
\hspace{-6mm}\includegraphics[width=7.8cm,clip]{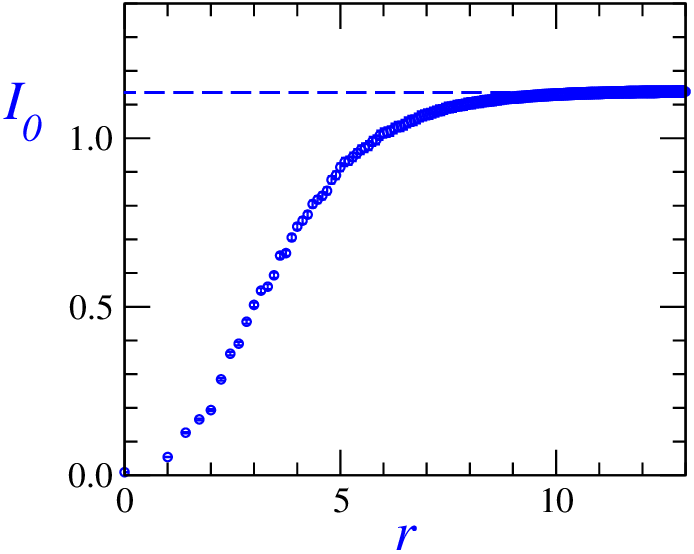}~\hspace{8mm}\includegraphics[width=7.8cm,clip]{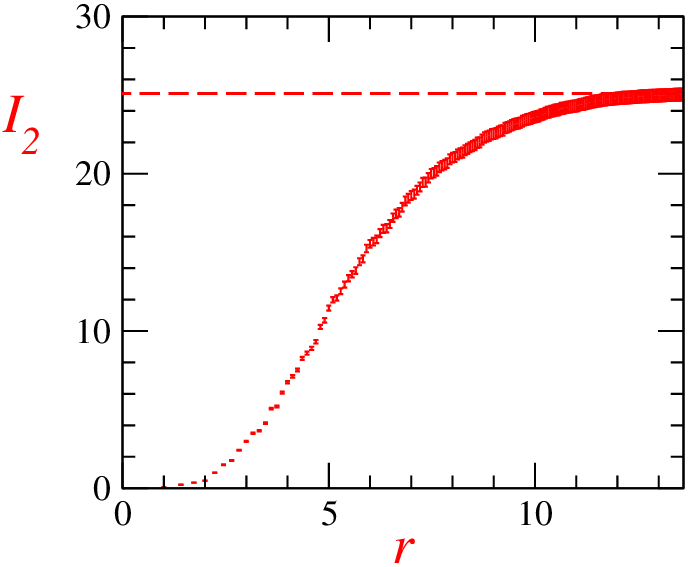}
\end{center}
\caption{\label{fig:3} \footnotesize \sl Summing over all distances: Plots show the saturation of the sum at larger distances of $I_0$ that corresponds to $-h_q/Z_A$, and of $I_2$ that is related to the second moment of the radial distribution.  Dashed lines are the final results for $I_{0,2}$ computed according to eq.~(\ref{integral8}). Illustration is provided for the lattices of Set 1, with $\kappa_{\rm sea}\equiv \kappa_q=0.1374$, and by using the HYP-2$^2$ static heavy quark propagator. Distances are given in lattice units.}
\end{figure}

Concerning the distribution $f_{PAS}(r)$, we obtain it by fitting all our data as indicated in eq.~(\ref{fPAS}) on the same interval $-t_x=t_y=t\in [4,7]$, and for each value of $r$. The result is illustrated in fig.~\ref{fig:1} where we show both $f_{PAS}(r)$, and $r^2 f_{PAS}(r)$.  In practice we also check in each situation that the sum over all lattice points indeed reproduces the result obtained by using the radial distribution.  We then checked that 
\bea
A_+(\Delta^2) = 4 \pi \int_0^\infty dr\ r^2  f_{PAS}(r)\,,
\eea
coincides with the value obtained from the double ratio~(\ref{double}), even if with a slightly larger error.~\footnote{In doing so we of course accounted for the ${\cal O}(am_q)$ effect in the renormalization constant. } Like we discussed in ref.~\cite{ours2}, one might worry whether or not $4\pi$ obtained from the angular integration might be spoiled by the fact that we are working on the cubic lattice which, of course, is not spherically symmetric. Thanks to the fact that the radial distribution $f_{PAS}(r)$ is a fast decreasing function and becomes compatible with zero before reaching the lattice boundaries, the effect of the cubic lattice geometry is irrelevant. This is the case for the range of pion masses considered in this work. To illustrate that point we show that 
\bea\label{integral8}
I_{2k}= 4\pi \sum_{r=0}^{L/2}  r^{2(k+1)} f_{PAS}(r) \simeq  4\pi{\int_0^\infty dr\ r^{2(k+1)}}   f_{PAS}(r) \,,
\eea
for $k=0,1$, i.e. that for larger distances the sum becomes saturated and is a good approximation to the continuum integral. $I_2$ above is related to the second moment of the radial distribution $f_{PAS}(r)$ that we will discuss later on. Illustration for the saturation of both $I_0$ and $I_2$ on the specific example of the lattice data is shown in fig.~\ref{fig:3}.

Finally, we should comment on the values of $h_q$ given in tab.~\ref{tab:2} that is obtained in a way discussed in subsec.~\ref{q20}, namely as:
\bea\label{eq:11}
h_q  = A_+(\Delta_q^2) \  R_{\Delta_q} = A_+(\Delta_q^2) \ {  \displaystyle{ \int_0^\infty dr\ r^2  {\sin( \Delta_q r )\over \Delta_q  r} f_{PAS}(r)} \over \displaystyle{  \int_0^\infty dr\ r^2  f_{PAS}(r)}} \,,
\eea
This is quite a unique opportunity in hadronic physics that we can check for the moment expansion. To that end we expand the integrand in eq.~(\ref{eq:9}) for small $q_z r$ and then by choosing $q_z=\Delta_q$ we have
\bea\label{eq:91}
&& \hspace*{-8mm}A_+(0)= 4\pi \left[ \int_0^\infty dr\ r^2  f_{PAS}(r)  - {1\over 6}{\Delta_q^2}\int_0^\infty dr\ r^4  f_{PAS}(r)  + {1\over 120}{\Delta_q^4}\int_0^\infty dr\ r^6  f_{PAS}(r) +\dots \right] \nn\\
&& \hfill \nn\\
&& \hspace*{5mm} =  A_+(\Delta_q^2) \left(1 - {\Delta_q^2\over 6} \left< r^2\right> + {\Delta_q^4\over 120} \left< r^4\right> + \dots \right)\,,
\eea
and after keeping the first couple of moments we see that 
\bea\label{eq:92}
 R_{\Delta_q} \simeq R_4 =  1 - {\Delta_q^2\over 6} \left< r^2\right> + {\Delta_q^4\over 120} \left< r^4\right> \,,
\eea
where the moments are 
\bea\label{eq:9}
 \left< r^{2k}\right> ={{\displaystyle{\int_0^\infty dr\ r^{2(k+1)}}}  f_{PAS}(r)\over {\displaystyle{ \int_0^\infty dr\ r^2  f_{PAS}(r) }} }\,.
\eea
We computed the moments $\left< r^{2}\right>$ and $\left< r^{4}\right>$ and observe that the first moment alone saturates $R_{\Delta_q}$ to about $80\%$, while $R_4$ roughly coincides with  $R_{\Delta_q}$. The table of results for the moments, as well as a comparison $R_4/R_{\Delta_q}$ is provided in Appendix of this paper. 
\subsection{Chiral extrapolation}
\begin{figure}
\psfrag{X00}{\large \sc Set 1 }
\psfrag{X23}{\large \sc Set 2 \& 3 }
\psfrag{AA2}{\LARGE ${\color{red} h_{q}}$}
\psfrag{BB2}{\LARGE ${\color{black}\hat g_{q}}$}
\psfrag{CC2}{\LARGE\!\!\!\!\! ${\color{blue}- \widetilde g_{q}}$}
\begin{center}\includegraphics[width=10.1cm,clip]{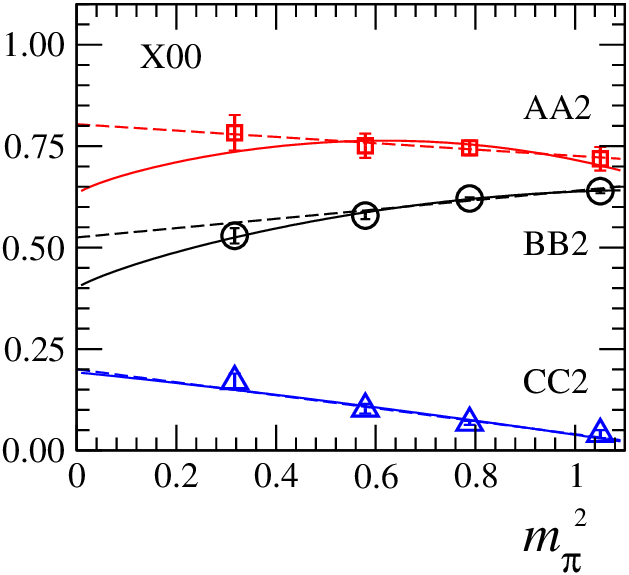}\\
\hfill \\
\vspace{11mm}
\includegraphics[width=10.1cm,clip]{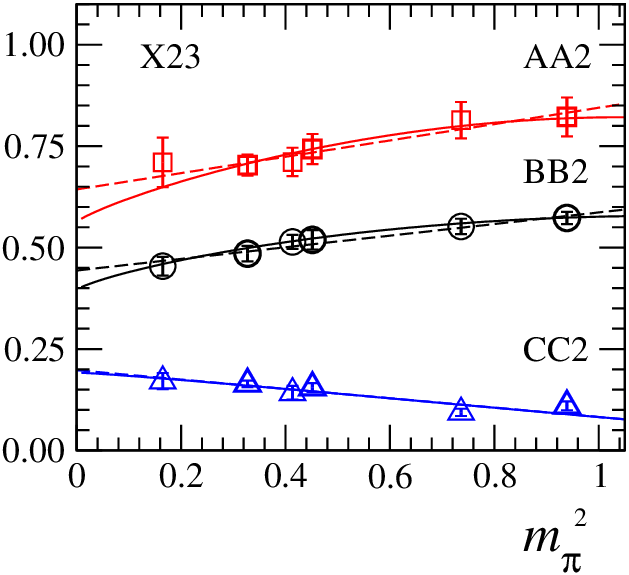}
\end{center}
\caption{\label{fig:4} \footnotesize \sl Chiral extrapolations: Dashed lines denote the linear extrapolations, and the full curves correspond to the results obtained by fitting the data to the expressions derived in HMChPT, c.f. eq.~(\ref{fit:hmchpt}). In the lower plot we combined the results from Set 2 (lighter) and Set 3 (thicker symbols). Results of the fit are given in tab.~\ref{tab:4}.} 
\end{figure}

To get the physically interesting couplings $h$, $g$ and $\widetilde g$ we need to perform a chiral extrapolation of the results listed in tab.~\ref{tab:2}. 
Since our results are obtained for the pion masses not too close to the chiral limit and exhibit a rather clean linear dependence in $m_q\propto m_\pi^2$, we first attempt a linear chiral extrapolation. 
In particular,
\bea\label{fit:linear}
&&\hat g_q= \hat g   \left( 1 + c_g  m_\pi^2\right)\,,\qquad \widetilde g_q= \widetilde g   \left( 1 + \widetilde c_g  m_\pi^2\right)\,,\nn\\
&&\hfill \nn\\
&& \qquad  \qquad h_q = h \left( 1 + c_h  m_\pi^2\right)\,.
\eea
Another possibility is to use the expressions derived in HMChPT~\cite{jernej}, in which the effects of the nearest orbital excitations has been taken into account. Those corrections read:
\bea\label{fit:hmchpt}
&&\hat g_q = g \left[ 1- {4 g^2\over (4 \pi f)^2} m_\pi^2\log m_\pi^2 - {h^2\over (4 \pi f)^2}{ m_\pi^2\over 8 \Delta_q^2} \left( 3 +{\widetilde  g\over g}\right) m_\pi^2\log m_\pi^2 + c_g m_\pi^2 \right]\,,\nn\\
&&\hfill \nn\\
&& \widetilde g_q = \widetilde g \left[ 1- {4 \widetilde g^2\over (4 \pi f)^2} m_\pi^2\log m_\pi^2 + {h^2\over (4 \pi f)^2}{ m_\pi^2\over 8 \Delta_q^2} \left( 3 +{  g\over \widetilde g}\right) m_\pi^2\log m_\pi^2 + \widetilde  c_g m_\pi^2 \right]\,,\nn\\
&&\hfill \nn\\
&& h_q = h \left[ 1- {3\over 4} { 3 g^2 + 3 \widetilde g^2 - 2 g \widetilde g \over (4 \pi f)^2} m_\pi^2\log m_\pi^2 - {h^2\over (4 \pi f)^2}{ m_\pi^2\over 2 \Delta_q^2} m_\pi^2\log m_\pi^2 + c_h m_\pi^2 \right]\,,
\eea
where, again, on the left hand side are the quantities we computed on the lattice (each carrying an index ``$q$") and on the right hand side are the desired quantities $h$, $g$ and $\widetilde g$.  The terms $\propto m_\pi^2/\Delta_q^2$ come from the inclusion of the heavy-light states of opposite parity in the chiral loops. Omitting those terms is equivalent to assuming that $m_\pi \ll \Delta$, which is the approximation needed to recover the usual HMChPT formulas for $\hat g$, and $\widetilde g$. To appreciate the difference, we will fit our data to the form in which the terms  $\propto m_\pi^2/\Delta_q^2$  are omitted and the corresponding results will be labelled as ``chi-1". The results obtained by using the full formulas~(\ref{fit:hmchpt}) will be denoted as ``chi-2".
The fits are illustrated in fig.~\ref{fig:4}, and the results are given in tab.~\ref{tab:4}. 

We did not combine all the data to make the final extrapolation. Instead we prefer to give our results separately for the data in which Iwasaki gauge action has been used (Set 1), and those that are obtained from the field configurations generated by using the Wilson plaquette gauge action (Set 2 and 3). Although the results are compatible, within the error bars, we note that  $h_q$ values in the Set 1 remains quite flat when the light quark mass is varied, while those obtained in Set 2 and 3 exhibit a positive slope in the light quark mass. 
\begin{table}[h!!]
\begin{sideways}\parbox{2.5cm}{ \hspace*{-4mm}{\color{blue}\large Set 1}}\end{sideways}~~\begin{tabular}{|c|c c c|c c c|} 
\hline
{\phantom{\huge{l}}} \raisebox{-.2cm} {\phantom{\huge{j}}}
{\sl Extrapolation} & $h$  & $\hat g$ &$ \widetilde g$ & $c_h$&   $c_g$ & $ \widetilde c_g$   \\ \hline \hline
{\phantom{\huge{l}}} \raisebox{-.2cm} {\phantom{\huge{j}}}
lin. [eq.~(\ref{fit:linear})]  &                              $0.80(2)$ & $0.53(3)$    & $-0.20(3)$  & $-0.10(2)$ & $0.22(8)$ & $-0.80(6)$	\\
{\phantom{\huge{l}}} \raisebox{-.2cm} {\phantom{\huge{j}}}
chi-1 [eq.~(\ref{fit:hmchpt})]                        & $0.69(2)$ & $0.43(1)$    & $-0.19(2)$  & $0.05(4)$ & $0.51(5)$ & $-0.80(7)$	\\
{\phantom{\huge{l}}} \raisebox{-.2cm} {\phantom{\huge{j}}}
chi-2 [eq.~(\ref{fit:hmchpt})]                         & $0.63(3)$ & $0.41(1)$    & $-0.19(2)$  & $0.13(8)$ & $0.59(3)$ & $-0.80(6)$	\\ \hline
\end{tabular} \\
\begin{sideways}\parbox{2.5cm}{ \hspace*{-11mm}{\color{blue}\large Set 2 and 3}}\end{sideways}~~\begin{tabular}{|c|c c c|c c c|} 
\hline
{\phantom{\huge{l}}} \raisebox{-.2cm} {\phantom{\huge{j}}}
{\sl Extrapolation}  & $h$  & $\hat g$ &$ \widetilde g$ & $c_h$&   $c_g$ & $ \widetilde c_g$   \\ \hline \hline
{\phantom{\huge{l}}} \raisebox{-.2cm} {\phantom{\huge{j}}}
lin. [eq.~(\ref{fit:linear})]  &                              $0.64(2)$ & $0.44(2)$    & $-0.20(2)$  & $0.31(6)$ & $0.39(9)$ & $-0.58(10)$	\\
{\phantom{\huge{l}}} \raisebox{-.2cm} {\phantom{\huge{j}}}
chi-1 [eq.~(\ref{fit:hmchpt})]                        & $0.58(2)$ & $0.40(1)$    & $-0.22(2)$  & $0.47(9)$ & $0.52(6)$ & $-0.57(11)$	\\
{\phantom{\huge{l}}} \raisebox{-.2cm} {\phantom{\huge{j}}}
chi-2 [eq.~(\ref{fit:hmchpt})]                         & $0.56(2)$ & $0.40(1)$    & $-0.19(2)$  & $0.47(10)$ & $0.53(5)$ & $-0.57(10)$	\\ \hline
\end{tabular}\caption{\label{tab:4}\footnotesize Results of the extrapolations of our results to the chiral limit.  }
\end{table}
As our final results we therefore present  $h$, $\hat g$, $\widetilde g$ separately for Set 1, and for Set 2\&3, namely:
\bea\label{results-final}
\text{ Set 1} &:& h=0.69(2)\left({}^{+11}_{-07}\right)\,,\quad \hat g=0.43(1)\left({}^{+10}_{-02}\right),\quad \widetilde g=-0.19(2)\left({}^{+1}_{-0}\right)\,,\nn\\
&&\hfill \nn\\
\text{ Set 2 \& 3} &:& h=0.58(2)\left({}^{+6}_{-2}\right)\,,\quad \hat g=0.40(1)\left({}^{+4}_{-0}\right),\quad \widetilde g=-0.22(2)\left({}^{+2}_{-0}\right)\,,
\eea
where as a central we take the values obtained from the extrapolation ``chi-1", and the difference with the results obtained through the linear extrapolation and by using the full formulas from eq.~(\ref{fit:hmchpt}) are used as the systematic error estimate. We believe it is better to leave these errors as such, and let the lattice data decide the preferred extrapolation curve by investing into simulations at ever lower light quark masses. We see for example from the data in Set 2 and 3 that the results obtained from the chiral extrapolations ``chi-1" and ``chi-2" are practically indiscernible, which is a consequence of the fact that the pion masses explored through these simulations were lower than those in the Set 1, where the difference remains pronounced for the coupling $h$. 
 
The emerging pattern from our results is quite clear, $\widetilde g < \hat g < h$. Knowing that $h$ is rather large, its impact on the chiral extrapolation of phenomenologically relevant quantities computed on the lattice should be examined carefully. We will return to that question in a separate paper~\cite{workinprogress}. Here we will simply comment on the decay width for the $S$-wave pion emission in the decay of the scalar $D_0^\ast$ and $B_0^\ast$-mesons . If we neglect the $1/m_{c,b}$ corrections to $h$, then by using eq.~(\ref{eq:xxx}) we obtain
\bea
&&\Gamma(D_0^\ast \to D \pi^\pm) = \left[ 0.26(13)_{\rm Set\ 1},0.18(6)_{\rm Set\ 2\& 3}\right]~\gev\,,\nn\\
&&\Gamma(B_0^\ast \to B \pi^\pm) = \left[0.22(7)_{\rm Set\ 1},0.13(6)_{\rm Set\ 2\& 3}\right]~\gev\,,
\eea
where we took the physical masses available from PDG~\cite{PDG}, except for the unmeasured scalar $B_0^\ast$ whose mass we get by imposing $m_{B_0^\ast} - m_B = m_{B_1}-m_{B^\ast}$, to get $m_{B_0^\ast}=5.68$~GeV. A larger error in the charm sector comes from the larger error on the mass $m_{D_0^\ast}=2.32(3)$~GeV, and the results can be compared to the PDG average of the experimentally established width of the scalar meson, $\Gamma(D_0^{\ast 0})=267(40)$~MeV~\cite{PDG}. In the case of $B_0^\ast$-decay our results agree with ref.~\cite{michael}.

Before concluding we should also comment on the comparison between our results~(\ref{results-final}) and the values obtained by using the light cone QCD sum rules (LCSR)~\cite{g-lcsr} summarized in ref.~\cite{Colangelo:1997rp} in which also $h$ has been computed:~\footnote{Since the overall sign of the couplings discussed here is convention dependent, we changed the signs of the results reported in ref.~\cite{Colangelo:1997rp} to be consistent with ours.} 
\bea\label{results-lcsr}
\text{ LCSR} &:& h=0.60\pm 0.13\,,\quad \hat g=0.17\pm 0.06,\quad \widetilde g=-0.10\pm 0.02\,.
\eea
Other QCD sum rule calculations lead to similar results~\cite{qsr-2} from which we see that the LCSR values for $\hat g$ and $\widetilde g$ are by about a factor of two smaller than those obtained on the lattice. This problem persists in the case of the charmed heavy quark, as discussed in ref.~\cite{dirac}, while the lattice QCD results of $g_{D^\ast D\pi}$ are consistent with the one extracted from the experimentally measured $\Gamma(D^{\ast \pm})$~\cite{CLEO}. In contrast to that situation, the value of the coupling $h$ obtained from LCSR agrees quite well with the lattice results.   
Finally, the results obtained in two classes of quark models discussed in ref.~\cite{ours3} also agree with the couplings presented here.

\section{Summary and conclusions\label{SecZ}}
In this paper we discussed the computation of the coupling of two heavy-light mesons to a soft pion in the static heavy quark limit. We focussed on the heavy-light mesons belonging to the $j^P=(1/2)^-$ [$L=0$], and $j^P=(1/2)^+$ [$L=1$] doublets. To compute the coupling that relates the heavy-light mesons belonging to two different doublets, $h$, one encounters a difficulty in getting to the point in which the four-momentum of the associated pion is $q^2\to 0$. We solve that difficulty by computing the radial distribution of the matrix element of the light quark axial current between the scalar and pseudoscalar mesons, which then allowed us to compute the suppression factor needed to convert the form factor $A_+(\Delta_q^2)$ to $h_q=A_+(0)$. In addition to this coupling we also computed the axial couplings that involve the heavy-light mesons belonging to the same doublet, namely $\hat g$ for $(1/2)^-$ states, and  $ \widetilde g$ for the $(1/2)^+$ ones. 
With all three couplings we were able to check on the chiral extrapolation in which the impact of the nearest excitations has been taken into account. All results presented in this paper are obtained from the correlation functions computed on the publicly available gauge field configurations obtained with $N_{\rm f}=2$ dynamical light quark flavors and by using the Wilson-Clover quark action and by using the Iwasaki (Set 1) and Wilson plaquette gauge actions (Set 2 \& 3). We implemented the improvement of the spectator static quark propagator (Wilson line), and from the data in which the Iwasaki action has been used we obtain:
\bea
\text{ Set 1} &:& h=0.71(2)(9)\,,\quad \hat g=0.47(1)(6),\quad \widetilde g=-0.19(2)(1)\,,
\eea
where we symmetrize the systematic error that comes from the chiral extrapolation. From the data obtained by using the Wilson plaquette gauge action, the results we get after combining the values of the couplings computed at two fine lattices, read:
\bea
\text{ Set 2 \& 3} &:& h=0.60(2)(4)\,,\quad \hat g=0.42(1)(2),\quad \widetilde g=-0.23(2)(1)\,.
\eea
Impact of the results of the present paper on the chiral extrapolation of the phenomenologically interesting quantities involving heavy-light mesons as computed on the lattice, will be addressed in a separate paper~\cite{workinprogress}.

\vspace*{2cm}

\section*{Acknowledgements}
We thank  the QCDSF and CP-PACS collaborations for making their gauge field configurations publicly available, the {\sl Centre de Calcul de l'IN2P3 \`a Lyon} for giving us access to their computing facilities, and ANR (contract ÒDIAMÓ ANR-07-JCJC-0031) for a partial support. E.C. also acknowledges the support from FEDER and MEC (Spain)  by Contract No.FIS2008-01661.

\newpage
\section*{Appendix}
As we mentioned in the text, thanks to the fact that we were able to compute the radial distribution of the desired matrix element, we were also able to check on the moment expansion. To do se we used eqs.~(\ref{eq:91}, \ref{eq:92}, \ref{eq:9}), and obtained the results listed in tab.~\ref{tab:3}. All results are given in lattice units.

\begin{table}[h!!]
\begin{tabular}{|c|c|c|c|c|c|c|} 
\hline
{\phantom{\huge{l}}} \raisebox{-.2cm} {\phantom{\huge{j}}}
$\beta\hspace*{5mm}$ & $\hspace*{5mm}\kappa_{q}\hspace*{5mm}$ & $\Delta_q$ & $\hspace*{4mm}R_\Delta\hspace*{4mm}$ &  $\hspace*{6mm}\langle r^2\rangle_q \hspace*{6mm}$  & $\hspace*{6mm}\langle r^4\rangle_q \hspace*{6mm}$ & $\hspace*{2mm} (R_4/R_\Delta)_q [\%]\hspace*{6mm}$ \\ \hline \hline
{\phantom{\huge{l}}} \raisebox{-.2cm} {\phantom{\huge{j}}}
$2.1$ &                                  $0.1357$ &  $0.269(4)$ & $0.889(26)$       & $17(1)$ & $606(26)$ & $93(4)$	\\
{\phantom{\huge{l}}} \raisebox{-.2cm} {\phantom{\huge{j}}}
{\color{blue}Set 1}             & $0.1367$ &  $0.273(4)$& $0.896(15)$      & $20(1)$ & $891(27)$ & $90(3)$	\\
{\phantom{\huge{l}}} \raisebox{-.2cm} {\phantom{\huge{j}}}
                                              & $0.1374$ &  $0.262(6)$ & $0.842(24)$     & $22(1)$ & $1127(65)$ & $93(5)$	\\
{\phantom{\huge{l}}} \raisebox{-.2cm} {\phantom{\huge{j}}}
                                              & $0.1382$ &  $0.272(6)$& $0.808(32)$    & $28(1)$ & $1835(108)$ & $92(7)$	\\
                                              \hline \hline 
{\phantom{\huge{l}}} \raisebox{-.2cm} {\phantom{\huge{j}}}
$5.29$ &                              $0.1355$ &  $0.232(8)$    & $0.829(32)$    & $30(1)$ & $1976(100)$ & $98(8)$	\\
{\phantom{\huge{l}}} \raisebox{-.2cm} {\phantom{\huge{j}}}
  {\color{blue}Set 2}           & $0.1359$ &  $0.210(5)$& $0.797(28)$      & $33(2)$ & $2542(185)$ & $99(6)$	\\
{\phantom{\huge{l}}} \raisebox{-.2cm} {\phantom{\huge{j}}}
                                              & $0.1362$&  $0.223(12)$ &$0.772(47)$    & $33(2)$ & $2502(177)$ & $101(13)$	\\
	 \hline \hline
{\phantom{\huge{l}}} \raisebox{-.2cm} {\phantom{\huge{j}}}
$5.40$ &                             $0.1356$ &  $0.216(6)$ & $0.840(35)$    & $31(1)$ & $2033(100)$ & $95(7)$	\\
{\phantom{\huge{l}}} \raisebox{-.2cm} {\phantom{\huge{j}}}
  {\color{blue}Set 3}           & $0.1361$ &  $0.233(7)$& $0.822(29)$     & $35(1)$ & $2609(149)$ & $91(8)$	\\
{\phantom{\huge{l}}} \raisebox{-.2cm} {\phantom{\huge{j}}}
           &                                    $0.13625$ &  $0.207(8)$ & $0.810(21)$     & $39(2)$ & $3249(180)$ & $95(10)$	\\
	 \hline 
\end{tabular}\caption{\label{tab:3}\footnotesize The values of the first couple of moments as obtained from our lattices are given in lattice units. The comparison of the full expression $R_\Delta$ and the one obtained by combining the moments $R_4$, as indicated in eq.~(\ref{eq:92}), is presented in the last column. For convenience we also give the values of the orbital splitting, i.e. $\Delta_q =  \widetilde {\cal E}_q - {\cal E}_q$ which is practically the mass difference between the scalar and pseudoscalar static-light mesons. Three Sets of data corresponds to parameters listed in tab.~\ref{tab:1}. }
\end{table}


\newpage 

\begin{thebibliography}{99}

\bibitem{g-ukqcd}
G.~M.~de Divitiis {\it et al.} [UKQCD Collaboration],
JHEP {\bf 9810} (1998) 010
[hep-lat/9807032]; 
%%CITATION = HEP-LAT 9807032;%%


\bibitem{g-orsay}
A.~Abada {\it et al.},
JHEP {\bf 0402} (2004) 016
[hep-lat/0310050].
%%CITATION = HEP-LAT 0310050;%%
%%CITATION = HEP-PH 9901431;%%

\bibitem{g-jlqcd}
  H.~Ohki, H.~Matsufuru and T.~Onogi,
  %``Determination of B*B pi coupling in unquenched QCD,''
  Phys.\ Rev.\  D {\bf 77} (2008) 094509
  [arXiv:0802.1563 [hep-lat]].
  %%CITATION = PHRVA,D77,094509;%%
  
\bibitem{ours1}
  D.~Becirevic, B.~Blossier, E.~Chang and B.~Haas,
  %``g(B*Bpi)-coupling in the static heavy quark limit,''
  Phys.\ Lett.\ B {\bf 679} (2009) 231
  [arXiv:0905.3355 [hep-ph]].
  %%CITATION = ARXIV:0905.3355;%%
%\cite{Bulava:2010ej}

\bibitem{g-alpha}
  J.~Bulava, M.~A.~Donnellan and R.~Sommer  [ALPHA Collaboration],
  %``The B*Bpi Coupling in the Static Limit,''
  PoS {\bf LATTICE2010} (2010) 303
  [arXiv:1011.4393 [hep-lat]].
  %%CITATION = POSCI,LATTICE2010,303;%%
  
%\cite{Detmold:2011bp}
\bibitem{g-detmold}
  W.~Detmold, C.~J.~Lin and S.~Meinel,
  %``Axial couplings and strong decay widths of heavy hadrons,''
  arXiv:1109.2480 [hep-lat].
  %%CITATION = ARXIV:1109.2480;%%

\bibitem{gDDpi1}
A.~Abada {\it et al.},
Phys.\ Rev.\ D {\bf 66} (2002) 074504
[hep-ph/0206237].
%%CITATION = HEP-PH 0206237;%%

\bibitem{gDDpi2}
  D.~Becirevic and B.~Haas,
  %``D* ---> D pi and D* ---> D gamma decays: Axial coupling and Magnetic moment
  %of D* meson,''
  Eur.\ Phys.\ J.\  C {\bf 71} (2011) 1734
  [arXiv:0903.2407 [hep-lat]].
  %%CITATION = EPHJA,C71,1734;%%


%\cite{Becirevic:1999fr}
\bibitem{dirac}
  D.~Becirevic and A.~L.~Yaouanc,
  %``g-hat coupling (g(B* B pi), g(D* D pi)): A quark model with Dirac
  %equation,''
  JHEP {\bf 9903} (1999) 021
  [arXiv:hep-ph/9901431];
  %%CITATION = JHEPA,9903,021;%%
  D.~Becirevic, J.~Charles, A.~LeYaouanc, L.~Oliver, O.~Pene and J.~C.~Raynal,
  %``Possible explanation of the discrepancy of the light-cone QCD sum rule
  %calculation of g(D* D pi) coupling with experiment,''
  JHEP {\bf 0301} (2003) 009
  [arXiv:hep-ph/0212177].
  %%CITATION = JHEPA,0301,009;%%


\bibitem{CLEO}
A.~Anastassov {\it et al.}  [CLEO Collaboration],
Phys.\ Rev.\ D {\bf 65} (2002) 032003
[hep-ex/0108043].
%%CITATION = HEP-EX 0108043;%%



%\cite{McNeile:2004rf}
\bibitem{michael}
  C.~McNeile, C.~Michael and G.~Thompson  [UKQCD Collaboration],
  %``Hadronic decay of a scalar B meson from the lattice,''
  Phys.\ Rev.\  D {\bf 70} (2004) 054501
  [arXiv:hep-lat/0404010].
  %%CITATION = PHRVA,D70,054501;%%


\bibitem{ours2}
  D.~Becirevic, E.~Chang and A.~L.~Yaouanc,
  %``On internal structure of the heavy-light mesons,''
  Phys.\ Rev.\  D {\bf 80} (2009) 034504
  [arXiv:0905.3352 [hep-lat]].
  %%CITATION = PHRVA,D80,034504;%%



\bibitem{green}
  A.~M.~Green, J.~Koponen, P.~Pennanen and C.~Michael  [UKQCD Collaboration],
  %``The charge and matter radial distributions of heavy-light mesons
  %calculated on a lattice with dynamical fermions,''
  Eur.\ Phys.\ J.\  C {\bf 28} (2003) 79
  [arXiv:hep-lat/0206015];
  %%CITATION = EPHJA,C28,79;%%
  %``The charge and matter radial distributions of heavy-light mesons
  %calculated on a lattice,''
  Phys.\ Rev.\  D {\bf 65} (2002) 014512
  [arXiv:hep-lat/0105027].
  %%CITATION = PHRVA,D65,014512;%%

\bibitem{hasenfratz}
  A.~Hasenfratz and F.~Knechtli,
  %``Flavor symmetry and the static potential with hypercubic blocking,''
  Phys.\ Rev.\  D {\bf 64} (2001) 034504
  [arXiv:hep-lat/0103029].
  %%CITATION = PHRVA,D64,034504;%%

%\cite{Della Morte:2005yc}
\bibitem{actions-dellamorte}
  M.~Della Morte, A.~Shindler and R.~Sommer,
  %``On lattice actions for static quarks,''
  JHEP {\bf 0508} (2005) 051
  [arXiv:hep-lat/0506008].
  %%CITATION = JHEPA,0508,051;%%


%\cite{Becirevic:2011cj}
\bibitem{ours3}
  D.~Becirevic, E.~Chang, L.~Oliver, J.~C.~Raynal and A.~Le Yaouanc,
  %``Spatial distributions in static heavy-light mesons: a comparison of quark
  %models with lattice QCD,''
  Phys.\ Rev.\  D {\bf 84} (2011) 054507
  [arXiv:1103.4024 [hep-ph]].
  %%CITATION = PHRVA,D84,054507;%%




\bibitem{Casalbuoni}
R.~Casalbuoni, A.~Deandrea, N.~Di Bartolomeo, R.~Gatto, F.~Feruglio and G.~Nardulli,
Phys.\ Rept.\  {\bf 281} (1997) 145
[hep-ph/9605342];
%%CITATION = HEP-PH 9605342;%%
A.~F.~Falk and M.~E.~Luke,
  %``Strong decays of excited heavy mesons in chiral perturbation theory,''
  Phys.\ Lett.\ B {\bf 292} (1992) 119
  [hep-ph/9206241];
  %%CITATION = HEP-PH/9206241;%%
  D.~Becirevic, S.~Fajfer and J.~F.~Kamenik,
  %``Chiral behavior of the $B^0$ ($d$, $s^{)}$ - $\bar{B}^0$( $d$, $s^{)}$
  %mixing amplitude in the standard model and beyond,''
  JHEP {\bf 0706} (2007) 003
  [arXiv:hep-ph/0612224];
 %%CITATION = JHEPA,0706,003;%%
%%CITATION = POSCI,LAT2007,063;%%



\bibitem{Eichten-Hill}
E.~Eichten and B.~Hill,
Phys.\ Lett.\ B {\bf 234} (1990) 511.
%%CITATION = PHLTA,B234,511;%%

%\cite{DellaMorte:2003mn}
\bibitem{alphaSN}
  M.~Della Morte {\it et al.} 
                  [ALPHA Collaboration],
  %``Lattice HQET with exponentially improved statistical precision,''
  Phys.\ Lett.\  B {\bf 581} (2004) 93
  [Erratum-ibid.\  B {\bf 612} (2005) 313]
  [arXiv:hep-lat/0307021].
  %%CITATION = PHLTA,B581,93;%%


\bibitem{Boyle}
P.~Boyle  [UKQCD Collaboration],
J.\ Comput.\ Phys.\  {\bf 179} (2002) 349
[hep-lat/9903033].
%%CITATION = HEP-LAT 9903033;%%


\bibitem{cp-pacs}
  A.~Ali Khan {\it et al.}  [CP-PACS Collaboration],
  %``Light Hadron Spectroscopy with Two Flavors of Dynamical Quarks on the
  %Lattice,''
  Phys.\ Rev.\  D {\bf 65} (2002) 054505
  [Erratum-ibid.\  D {\bf 67} (2003) 059901]
  [arXiv:hep-lat/0105015].
  %%CITATION = PHRVA,D65,054505;%%

\bibitem{qcdsf}
A.~Ali Khan {\it et al.}  [QCDSF Collaboration],
  %``Accelerating the hybrid Monte Carlo algorithm,''
  Phys.\ Lett.\ B {\bf 564} (2003) 235
  [hep-lat/0303026];
  %%CITATION = HEP-LAT/0303026;%%
 A.~A.~Khan {\it et al.},
  %``Axial coupling constant of the nucleon for two flavours of dynamical quarks in finite and infinite volume,''
  Phys.\ Rev.\ D {\bf 74} (2006) 094508
  [hep-lat/0603028].
  %%CITATION = HEP-LAT/0603028;%%


%\cite{Khan:2006de}
\bibitem{r047}
C.~Aubin {\it et al.},
  %``Light hadrons with improved staggered quarks: Approaching the continuum
  %limit,''
  Phys.\ Rev.\  D {\bf 70} (2004) 094505
  [arXiv:hep-lat/0402030]; 
  %%CITATION = PHRVA,D70,094505;%%
  A.~A.~Khan {\it et al.},
  %``Axial coupling constant of the nucleon for two flavours of dynamical quarks
  %in finite and infinite volume,''
  Phys.\ Rev.\  D {\bf 74} (2006) 094508
  [arXiv:hep-lat/0603028].
  %%CITATION = PHRVA,D74,094508;%%



  %\cite{Blossier:2009bx}
\bibitem{Blossier:2009bx}
  B.~Blossier {\it et al.}  [ETM Collaboration],
  %``Pseudoscalar decay constants of kaon and D-mesons from Nf=2 twisted mass
  %Lattice QCD,''
  JHEP {\bf 0907} (2009) 043
  [arXiv:0904.0954 [hep-lat]].
  %%CITATION = JHEPA,0907,043;%%



%\cite{Lepage:1992xa}
\bibitem{TI}
  G.~P.~Lepage and P.~B.~Mackenzie,
  %``On the viability of lattice perturbation theory,''
  Phys.\ Rev.\  D {\bf 48} (1993) 2250
  [arXiv:hep-lat/9209022].
  %%CITATION = PHRVA,D48,2250;%%


 
%\cite{Gockeler:2010yr}
\bibitem{Gockeler:2010yr}
  M.~Gockeler, R.~Horsley, Y.~Nakamura, H.~Perlt, D.~Pleiter, P.~E.~L.~Rakow, A.~Schafer and G.~Schierholz {\it et al.},
  %``Perturbative and Nonperturbative Renormalization in Lattice QCD,''
  Phys.\ Rev.\ D {\bf 82} (2010) 114511
  [arXiv:1003.5756 [hep-lat]].
  %%CITATION = ARXIV:1003.5756;%%



\bibitem{jernej}
S.~Fajfer and J.~F.~Kamenik,
  %``Chiral loop corrections to strong decays of positive and negative  parity
  %charmed mesons,''
  Phys.\ Rev.\  D {\bf 74} (2006) 074023
  [arXiv:hep-ph/0606278].
  %%CITATION = PHRVA,D74,074023;%%


\bibitem{workinprogress} 
D.~Becirevic, E.~Chang, S.~Fajfer, J.~Kamenik, F.~Sanfilippo, {\it in preparation}. 



%\cite{Belyaev:1994zk}
\bibitem{g-lcsr}
  V.~M.~Belyaev, V.~M.~Braun, A.~Khodjamirian and R.~Ruckl,
  %``D* D pi and B* B pi couplings in QCD,''
  Phys.\ Rev.\  D {\bf 51} (1995) 6177
  [arXiv:hep-ph/9410280];
  %%CITATION = PHRVA,D51,6177;%%
  A.~Khodjamirian, R.~Ruckl, S.~Weinzierl and O.~I.~Yakovlev,
  %``Perturbative QCD correction to the light cone sum rule for the B* B pi and D* D pi couplings,''
  Phys.\ Lett.\ B {\bf 457} (1999) 245
  [hep-ph/9903421];
  %%CITATION = HEP-PH/9903421;%%
P.~Ball and R.~Zwicky,
  %``New results on B ---> pi, K, eta decay formfactors from light-cone sum rules,''
  Phys.\ Rev.\ D {\bf 71} (2005) 014015
  [hep-ph/0406232].
  %%CITATION = HEP-PH/0406232;%%
  
  
%\cite{Colangelo:1997rp}
\bibitem{Colangelo:1997rp}
  P.~Colangelo and F.~De Fazio,
  %``QCD interactions of heavy mesons with pions by light cone sum rules,''
  Eur.\ Phys.\ J.\  C {\bf 4}, 503 (1998)
  [arXiv:hep-ph/9706271].
  %%CITATION = EPHJA,C4,503;%%



\bibitem{qsr-2}
  P.~Colangelo {\it et al.},
  %``Strong coupling of excited heavy mesons,''
  Phys.\ Rev.\ D {\bf 52} (1995) 6422
  [hep-ph/9506207];
  %%CITATION = HEP-PH/9506207;%%
T.~M.~Aliev and M.~Savci,
  %``The Strong g(B)**B(pi) coupling constant in full QCD,''
  J.\ Phys.\ G G {\bf 22} (1996) 1759
  [hep-ph/9604258];
  %%CITATION = HEP-PH/9604258;%%
Y.~-B.~Dai and S.~-L.~Zhu,
  %``Couplings of pions with excited heavy mesons from light cone QCD sum rules in the leading order of HQET,''
  Eur.\ Phys.\ J.\ C {\bf 6} (1999) 307
  [hep-ph/9802227].
  %%CITATION = HEP-PH/9802227;%%



\bibitem{PDG}
  K.~Nakamura {\it et al.}  [Particle Data Group],
  %``Review of particle physics,''
  J.\ Phys.\ G {\bf 37} (2010) 075021.
  %%CITATION = JPHGB,G37,075021;%%


\end{thebibliography}
\end{document}